\documentclass[12pt]{article}
\usepackage{amsmath,amssymb,amsthm}
\usepackage{graphicx}                        
\usepackage{caption}                        
\usepackage{float}                        
\usepackage{ucs}                        
\usepackage[utf8x]{inputenc}            
\usepackage[russian,english]{babel}            
\title{\vskip -1.0 cm \bf Light    simulation in 
plastic scintillator  strip 
with  embedded  wavelength shifting fiber\vskip 1.0cm}
\author{ \bf  \Large  Usubov\,Z.U.      \\
\\
Joint Institute for Nuclear Research, Dubna, Russia    }
\theoremstyle{plain}

\begin{document}

\maketitle
\setlength{\parindent}{0pt}
\begin{abstract}
\noindent

The simulation study of the 
  light yield and attenuation in the plastic scintillator was performed.
The  wavelength shifting fiber readout
was embedded 
in the grooves machined  along the entire
  strip surface. 
  The scintillator strips was  irradiated
   with a radiation source ${}^{90}Sr$  or cosmic muons along and across
the strip.

\end{abstract}
%%\large  
\large  {
\section{Introduction}
$\quad$Cosmic muons are the  
important contributors to background processes when 
search    for the  conversion of a muon to an  electron\cite{come,mu2e}. Cosmic
ray veto geometry surrounding the detectors and stopping target should
be carefully eliminated this background. Passive and active  shielding
should provide background of $\sim$0.1 events in signal window for 3 year run of
experiments. Scintillator-based active shielding will consist of four layers.
The scintillator strips will read out by multiclad  wavelength
shifting (WLS) fibers connected to photodetectors. Most promising devices 
based on new technologies  used as photodetectors are the  silicon photomultipliers 
(SiPM)\cite{golo}. Photon  detection efficiency of SiPM is  dependent on wavelength of
the photon, temperature, dark count\cite{yang} etc., and very difficult to be
measured and even more difficult to model\cite{kjha}. For this reason, we are
limited only to the study of the light propagation and collection.     
%%% and our simulation data have mainly comparative importance. 
Even for
this purpose, a  Monte Carlo simulation can adequately predict the 
experimental results only if the detector parameters are 
sufficiently close to their true values. Some of parameter, e.g. surface
boundaries descriptions, can be tuned by using measurements for   
particular scintillator strip configurations.
%%%\clearpage{\pagestyle{empty}\cleardoublepage}
%%\pagestyle{empty}
\begin{figure}[htp]                             %%   [H]      
%%%%\vskip -0.0cm \hskip  2.2cm \includegraphics[scale=0.6]{Dec_spec.png}
\begin{center}    
\includegraphics[   width=0.6\textwidth, height=0.4\textwidth]{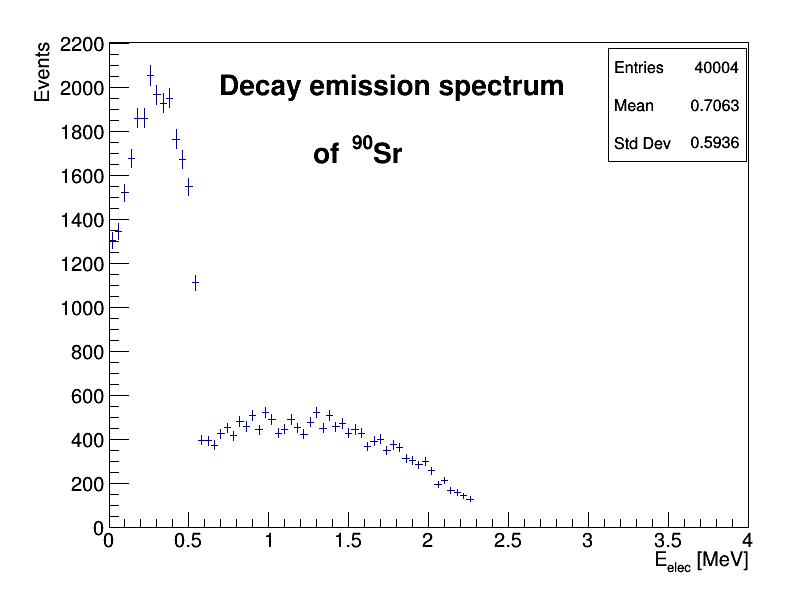}
\captionof{figure}{ The $\beta$-particle spectrum of  ${}^{90}Sr$ provided by Geant4 
simulation.}
\label{Norm}
 
\end{center}
\end{figure}

\section{ Some features of the plastic scintillator  strip modeling }
$\quad$The Monte Carlo 
simulation  with all possible processes play a crucial 
 role in the feasibility study of the proposed detector module and in identifying
detector parameter values. Low-energy optical photons (photons with a
wavelength much greater than the typical atomic spacing) undergo the
following processes: bulk absorption, Rayleigh scattering, reflection
and refraction at medium boundaries, and wavelength shifting.
%%%\clearpage{\pagestyle{empty}\cleardoublepage}
%%\pagestyle{empty}
\begin{figure}[htp]                             %%   [H]      
%%%%\vskip -0.0cm \hskip  2.2cm \includegraphics[scale=0.6]{Dec_spec.png}
\begin{center}    
\includegraphics[   width=0.6\textwidth, height=0.4\textwidth]{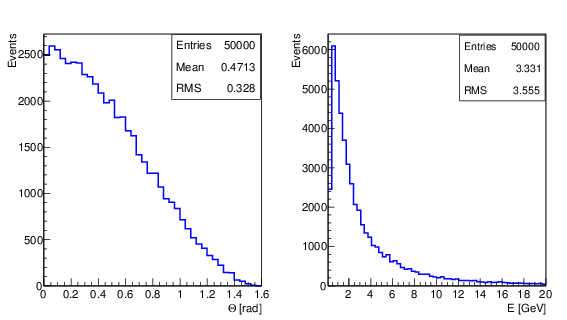}
\captionof{figure}{ The zenith  angle and energy distribution for simulated         
cosmic muons.}
\label{Norm}
 
\end{center}
\end{figure}

\quad The boundary processes on all scintillator play an important role in  tracing
 photons in strips. Compared to   them, photon self-absorption  in 
scintillator is less significant\cite{knoll}. In Geant4.10.06 \cite{gea4} 
 simulation we combined 
the polished scintillator surface finishes with the backpainted wrapping
option which represents diffuse (Lambertian) reflection. In this
simulation we use the UNIFIED model for the processes between two
dielectric materials.
%%%\clearpage{\pagestyle{empty}\cleardoublepage}
%%\pagestyle{empty}
\begin{figure}[htp]                             %%   [H]      
%%%%\vskip -0.0cm \hskip  2.2cm \includegraphics[scale=0.6]{Dec_spec.png}
\begin{center}    
\includegraphics[ width=0.6\textwidth, height=0.4\textwidth]{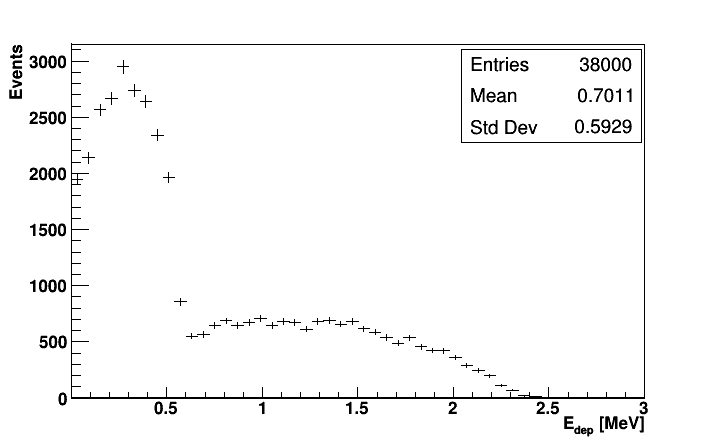}
\captionof{figure}{ Energy deposited  in scintillator strip with dimension 4*1*300\,$cm^{3}$ 
and ${}^{90}Sr$ radiation source (see the text).}
\label{Norm}
 
\end{center}
\end{figure}

\section{Simulation and results}

$\quad$ The peak of emission light of a plastic scintillator 
(e.g., Saint Gobain BC400 series) does not matches the peak sensitivity of used 
photodetectors. To solve this problem it is necessary to use the WLS fiber to
transfer light to the photodetector. In this simulation the fiber are multiclad
consisting of a scintillating core surrounded by an acrylic inner cladding
%%%\clearpage{\pagestyle{empty}\cleardoublepage}
%%\pagestyle{empty}
\begin{figure}[htp]                             %%   [H]      
%%%%\vskip -0.0cm \hskip  2.2cm \includegraphics[scale=0.6]{Dec_spec.png}
\begin{center}    
\includegraphics[ width=0.6\textwidth, height=0.4\textwidth]{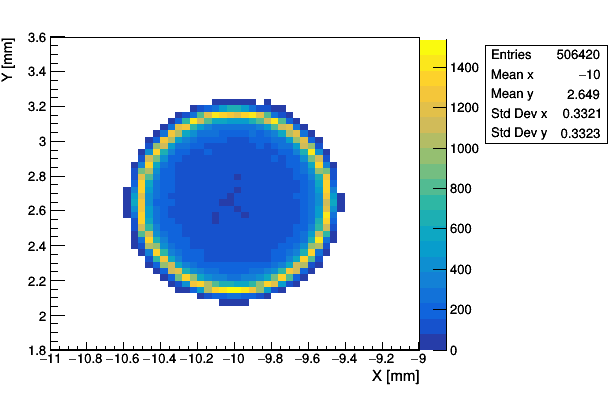}
\captionof{figure}{ Light intensity distribution in the fiber cross-section at the   
photodetector side of the scintillator strip.}
\label{Norm}
 
\end{center}
\end{figure}

and an outer cladding which made of a fluor-acrylic material (similar
to the Kuraray double clad fibers of type Y11(200)\cite{kura}). 
It was assumed that in a scintillator strip a mean value of 10000
scintillation photons per MeV of deposited energy  were emitted.
For this scintillator, the maximum emission is at a wavelength of 431\,nm 
and refractive index is 1.58.

\quad  For WLS fiber attenuation length of 500\,cm for its own radiation, and for
plastic scintillator attenuation length of 300\,cm are assumed.
The total diameter
of fiber is 1.2\,mm. The total thickness of  cladding structure is 6\%
of the diameter of a fiber. In this simulation the strip contains one or two
 co-extruded grooves with 3\,mm depth and 1.3\,mm width
 for insertion of the WLS fiber. The selected strip and fiber parameters
are close to those used in the test-beam measurements at JINR (Dubna, Russia). 
%%%\clearpage{\pagestyle{empty}\cleardoublepage}
%%\pagestyle{empty}
\begin{figure}[htp]                             %%   [H]      
%%%%\vskip -0.0cm \hskip  2.2cm \includegraphics[scale=0.6]{Dec_spec.png}
\begin{center}    
\includegraphics[   width=0.6\textwidth, height=0.4\textwidth]{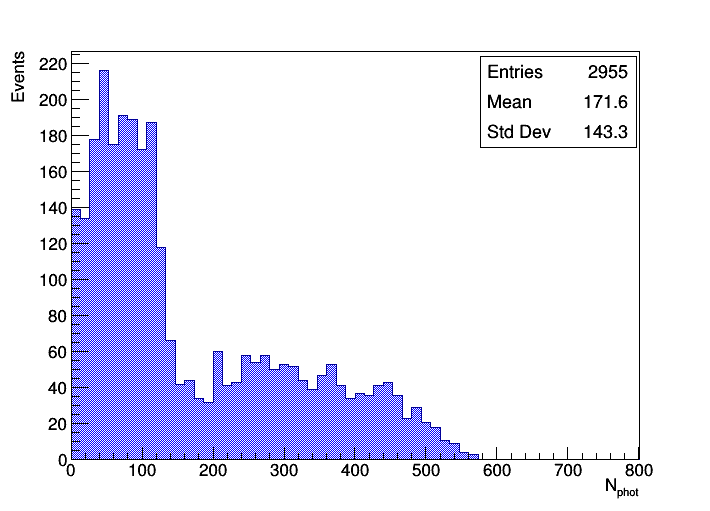}
\captionof{figure}{ The  distribution of the number of photons at the photodetector side when    
the middle of the strip irradiated by  ${}^{90}Sr$ source.}
\label{Norm}
 
\end{center}
\end{figure}

\quad  This simulation was performed using Geant4  for plastic
scintillator with the dimension 4*1*300\,cm${}^3$  and co-extruded TiO$_{2}$
white diffuse reflective (R=98$\%$) coating.
The strip contain one at the center or two grooves at a  
distance of  2\,cm from each other along the entire length of scintillator strip.
 We collect photons from a 
WLS fiber at one of the strip ends (hereinafter referred to as 
photodetector side). On the photodetector side at the fiber end
the photons are fully absorbed.  The opposite ends of the fibers are
blackened. 
%%%\clearpage{\pagestyle{empty}\cleardoublepage}
%%\pagestyle{empty}
\begin{figure}[htp]                             %%   [H]      
%%%%\vskip -0.0cm \hskip  2.2cm \includegraphics[scale=0.6]{Dec_spec.png}
\begin{center}    
\includegraphics[ width=0.6\textwidth, height=0.4\textwidth]{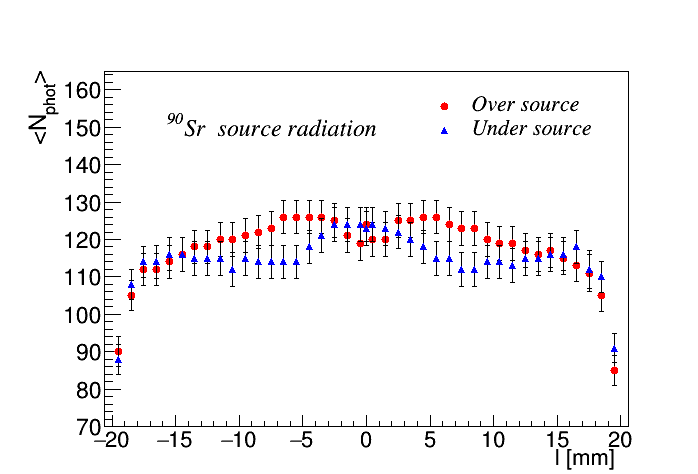}    
\captionof{figure}{ A comparison of light yield in scintillator strip with one       
fiber when ${}^{90}Sr$ source located   over and under the strip.}
\label{Norm}
 
\end{center}
\end{figure}

\quad The $^{90}Sr$   source was  simulated in the Geant4  framework.   
 The source provides an electron flux in a wide energy  range 
up to 2.3\,MeV (see Fig.1). The radiation source was enclosed in a shell with  
a lead collimator. The diameter of the collimator outlet was 1\,mm.
The source was located  at a distance of 2\,mm above or below the scintillator strip.

\quad Cosmic muons were generated according to\cite{volk}  in the range 
0.3-5000\,GeV. The zenith   angle and energy distribution for simulated cosmic
muons are displayed in Figure\,2.
%%%\clearpage{\pagestyle{empty}\cleardoublepage}
%%\pagestyle{empty}
\begin{figure}[htp]                             %%   [H]      
%%%%\vskip -0.0cm \hskip  2.2cm \includegraphics[scale=0.6]{Dec_spec.png}
\begin{center}    
\includegraphics[width=0.6\textwidth, height=0.4\textwidth]{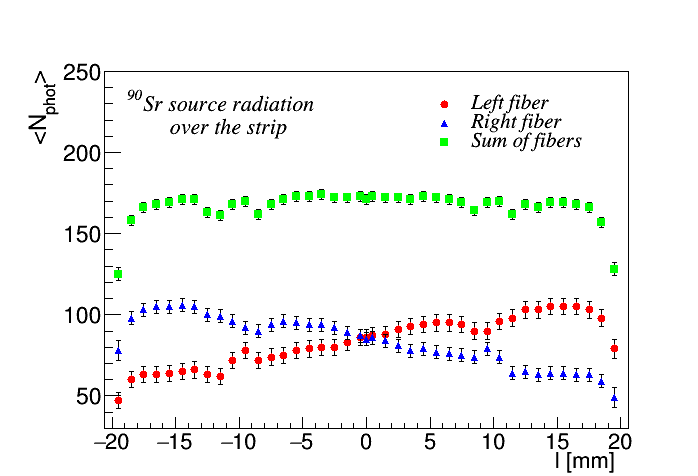}
\captionof{figure}{ A comparison of light yield in two fibers when ${}^{90}Sr$       
source is located over the scintillator strip.}
\label{Norm}
 
\end{center}
\end{figure}

%%%\clearpage{\pagestyle{empty}\cleardoublepage}
%%\pagestyle{empty}
\begin{figure}[htp]                             %%   [H]      
%%%%\vskip -0.0cm \hskip  2.2cm \includegraphics[scale=0.6]{Dec_spec.png}
\begin{center}    
\includegraphics[width=0.6\textwidth, height=0.4\textwidth]{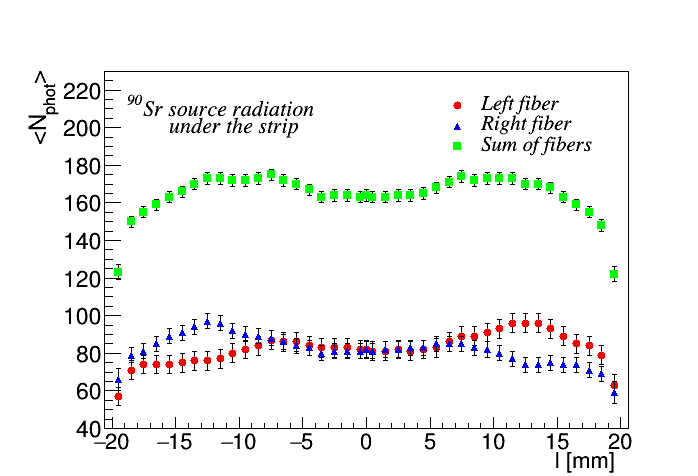}
\captionof{figure}{ A comparison of light yield in two fibers when ${}^{90}Sr$       
source is located under the scintillator strip.}
\label{Norm}
 
\end{center}
\end{figure}

\quad In Figure\,3, we show the distribution of the energy deposited in the scintillation
strip when the middle of the  side 4*300\,cm${}^{2}$ is irradiated with a ${}^{90}Sr$ source.

\quad Figure\,4 shows the light intensity distribution in the end  of a fiber
as seen by the photodetector side. This simulation study   show that
the light intensity increases towards the edge of the fiber core. The mean
wavelength of light collected by the photodetector is 535\,nm. 

\quad The distribution of the photon number at the photodetector side
when the middle of the strip side 4*300\,cm${}^{2}$ is
irradiated with  a ${}^{90}Sr$ is shown in Figure\,5.

\quad  Figure\,6 show the light yield when strip with one fiber irradiated
with a  $^{90}Sr$  source which located over or  under the strip at Z=0.0\,cm.
The strip is located at X=$\pm$\,2.0\,cm, Y=$\pm$\,0.5\,cm, Z=$\pm$\,150.0\,cm.       
The first point is 0.5\,mm away from the side (X=19.5\,mm)  and  each step is 
1\,mm (X-coordinate).
Points close to X=0.0\,mm are located at a distance of 0.5\,mm.
In this and the following Figures, 500 events were simulated for each point.

\quad  In Figures\,7  and 8,  the same thing  is shown as in Figure\,6  but for
the case with two fibers in the strip.  
Note that, in both cases (strip with one and two 
fibers), the behaviors of the light yield when a radiation source is above and
below the strip differ from each other. But this difference is not significant.

\quad  Figure\,9 shows the relation between the mean number of optical             
photons detected at the photodetector side and the distance between point
of impact of electrons from the radiation source and photodetector side.
This graph is fitted by a function
\vskip 0.5 cm
$\qquad\qquad\quad  N_{phot}(z) = A*e^{-z/\lambda_{1}} + B*e^{-z/\lambda_{2}}.$
\vskip 0.5 cm

Note that this formula was proposed by Kaiser et al.\cite{kais} for the case 
when light is collected from the ends of the scintillator using a photomultipliers.
The first term is the transmission behavior for photons that travel directly 
to the photodetector side.                                                          
%%%\clearpage{\pagestyle{empty}\cleardoublepage}
%%\pagestyle{empty}
\begin{figure}[htp]                             %%   [H]      
%%%%\vskip -0.0cm \hskip  2.2cm \includegraphics[scale=0.6]{Dec_spec.png}
\begin{center}    
\includegraphics[ width=0.6\textwidth, height=0.4\textwidth]{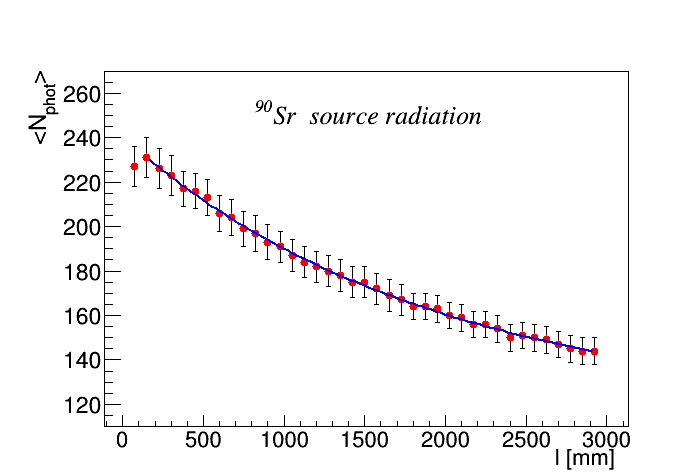}
\captionof{figure}{ The simulated light attenuation in a scintillator strip with     
two fiber with ${}^{90}Sr$ source irradiation. The first point is excluded from fit.}
\label{Norm}
\end{center}
\end{figure}

The second term is the transmission behavior for photons that hit the detector
after a series of reflection on a scintillator surface.
The first point in the Figure is 75 mm away from the photodetector side
and each step is 75\,mm. The curve in the figure corresponds to the parameters
 $\lambda_{1}$=43.5\,m and $\lambda_{2}$=1.75\,m  with almost 100\% errors.

$\quad$To study the light attenuation when the strip surface is  irradiated
by cosmic muons  we retreated on each side of the surface by 1\,mm  and 
divided it into 40 equal parts.   Each sector has been 
uniformly  irradiated by 500 muons with energies, azimuth and
zenith angles modeled accordingly to\cite{volk}. In Figure 10, we demonstrate
the light attenuation for this case. The points in the Figure are located in the
center of each of the 40 sections. For the given points, the results of one
exponential and double exponential fit are the same (blue curve in the Figure), 
$\lambda_{1}=\lambda_{2}$=5.88\,m. The green curve in  Figure corresponds
to the fit by the formula
\vskip 0.5 cm
$\qquad\qquad\quad  N_{phot}(z) = A*e^{-z/\lambda} + B,$
\vskip 0.5 cm
where $\lambda=$2.32\,m.
%%%\clearpage{\pagestyle{empty}\cleardoublepage}
%%\pagestyle{empty}
\begin{figure}[htp]                             %%   [H]      
%%%%\vskip -0.0cm \hskip  2.2cm \includegraphics[scale=0.6]{Dec_spec.png}
\begin{center}    
\includegraphics[ width=0.6\textwidth, height=0.4\textwidth]{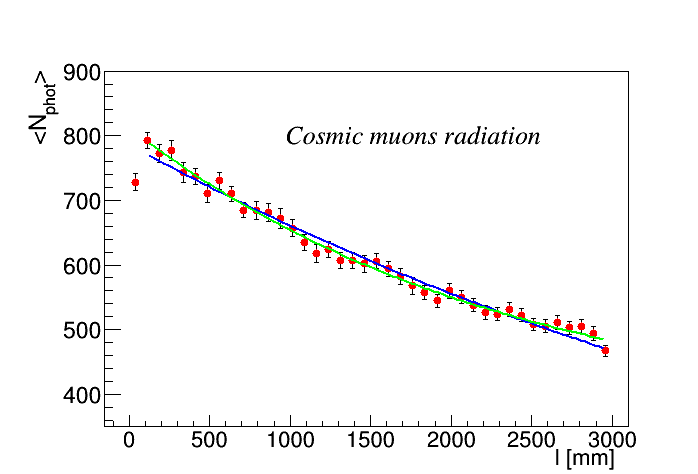}
\captionof{figure}{ The simulated light attenuation in a scintillator strip with     
cosmic muon irradiation. The first point is  excluded from fit.}
\label{Norm}
\end{center}
\end{figure}

\section{Conclusion}
\qquad In this note, we modelled  the  light output and attenuation in a scintilltion
strip with dimensions of  4*1*300\,cm${}^{3}$.                                 
The  simulated  radiation source ${}^{90}Sr$ and cosmic muons were used as beam particles.
The scintillation 
strip was irradiated both from the side of the embedded fibers and from
the opposite side along and across the strip.
Optical photons was collected
from one and two fibers embedded in the strip along the entire length. It
was shown that the attenuation of light depending on the distance to the
photodetector is described by a double exponential function.
%%%%\vspace {0.4cm}

$\quad$ We are sincerely grateful to  Z.\,Tsamalaidze and         
 Yu.\,Davydov for initiating this work.                             
}
\small{
\bibliographystyle{plain}
\bibliography{martingales}

\begin {thebibliography}{99}
\bibitem{come} 
K.\,Akhmetshin et al., Letter of Intend for Phase-I of the COMET Experiment
at J-PARC, KEK/J-PARC-PAC 2011-27, March 11, (2012).
\bibitem{mu2e}
L.\,Bartoszek et al., Mu2e Technical Design Report, FERMILAB-TM-2594, FERMILAB-DESIGN-2014-01,  
arXiv:1501.05241[physics.ins-det].
\bibitem{golo}
V.\,Golovin and V.\,Savelev, Nucl.Instrum.Meth. A442 (2000) 223; A518 (2004) 560.
\bibitem{yang}
S.K.\,Yang et al., Opt. Express 22 (2014) 216.         
\bibitem{kjha}
P.\,Eckert, R.\,Stamen, and H.\,Schultz-Coulon, JINST 7 (2012) P08011; \\
A.K.\, Jha et al., IEEE Trans.Nucl.Sci. 30(1) (2013) 336.    
\bibitem{knoll}
G.F.\,Knoll, Radiation detection and measurement, Third Edition, John Wiley and
Sons, Inc., Ann Arbor, (2000) 247.
\bibitem{gea4}
S.\,Agostinelli  et al., Geant4 - a simulation toolkit, 
Nucl.Instrum.Meth. A506 (2003) 250, \\
https://geant4.web.cern.ch/support/user\_documentation
\bibitem{kura}
KURARAY CO., LTD., http://www.kuraray.co.jp/en/.
\bibitem{volk}
L.N.\,Volkova, G.T.\,Zatsepin, L.A.\,Kuzmichev, Yad.Fiz. 29 (1979) 1252; Sov.J.Nucl.Phys 29 
(1979) 645.
\bibitem{kais}
S.C.\,Kaiser, J.A.M.\,de Villiers, IEEE Transactions on Nuclear Science, 1964, 11(3) 29.
\end{thebibliography}

\end{document}